\documentclass{article}

\usepackage{amssymb}
\usepackage{graphicx}
\usepackage{calc,listings,color,multirow}

\usepackage[ruled,nothing]{algorithm}
\usepackage{algorithmic}

\usepackage{url,xspace}

\newcommand{\linbox}{{\sc LinBox}\xspace}

\newcommand{\accumulatewhile}{ \textbf{accumulate\_until} }
\newcommand{\Accumulatewhile}{ \textbf{Accumulate\_until} }

\begin{document}

\title{\linbox founding scope allocation, parallel building blocks, and
  separate compilation}

\urldef\jgdemail\url{Jean-Guillaume.Dumas@imag.fr}
\urldef\cpemail\url{Clement.Pernet@imag.fr}
\urldef\bdsemail\url{saunders@udel.edu}
\urldef\tgemail\url{Thierry.Gautier@inrialpes.fr}

\author{Jean-Guillaume Dumas\thanks{Laboratoire J. Kuntzmann, Universit\'e de
  Grenoble. 51, rue des Math\'ematiques, umr CNRS 5224, bp 53X, F38041
  Grenoble, France, \jgdemail. Part of this work was done while the first author was visiting the Claude Shannon Institute of the University College Dublin, Ireland, under a CNRS grant.}
\and Thierry Gautier\thanks{Laboratoire LIG, Universit\'e de
  Grenoble. umr CNRS, F38330 Montbonnot, France. \tgemail, \cpemail. Part of this work was done while the second author was visiting the ArTeCS group of the  University Complutense, Madrid, Spain.}
\and Cl\'ement Pernet\footnotemark[2]
\and B. David Saunders\thanks{University of Delaware, Computer and
  Information Science Department. 
Newark / DE / 19716, USA. \bdsemail.}
}

\maketitle
\abstract{
To maximize efficiency in time and space, allocations and
deallocations, in the exact linear algebra library \linbox, must
always occur in the founding scope. This provides a simple lightweight
allocation model. We present this model and its usage for the
rebinding of matrices between different coefficient domains. We also
present automatic tools to speed-up the compilation of template
libraries and a software abstraction layer for the introduction of
transparent parallelism at the algorithmic level.
}

\section{Introduction}
As a building block for a wide range of applications, computational
exact linear algebra has to conciliate efficiency and genericity. The
goal of the \linbox project is to address this problem in the design
of an efficient general-purpose \texttt{C++} open-source library for
exact linear algebra over the integers, the rationals, and finite
fields.
Matrices can be either dense, sparse or black box (i.e. viewed as a linear
operator, acting on vectors only). The library proposes a set of high level
linear algebra solutions, such as the rank, the determinant, the solution of a
linear system, the Smith normal form, the echelon form, the characteristic
polynomial, etc. Each of these solutions involves a hybrid combination
of several specialized 
algorithms depending on the domain, and the type of matrix. Over a
finite field, 
the building blocks are an efficient implementation of Wiedemann and block
Wiedemann algorithms combined with preconditioners~\cite{CEKSTV:2002:EP} for
black box matrices, a sparse Gaussian elimination for sparse matrices and the
BLAS based dense linear algebra techniques of the \texttt{FFLAS}
library~\cite{DGP:2008:dlaff} for dense matrices. The solutions over
the integers 
and rationals are lifted from modular computations by a Chinese remainder
algorithm or $p$-adic lifting.
The design is based on high genericity to allow us to write efficient algorithms independent of the many 
representations of domains and matrices. As a middleware, the library relies on the
efficiency of kernel libraries such as  \texttt{GMP}\footnote{\url{gmplib.org,www-ljk.imag.fr/CASYS/LOGICIELS/givaro,www.shoup.net/ntl,math-atlas.sourceforge.net,sagemath.org,www.maplesoft.com.}},
\texttt{Givaro}\footnotemark[4],
\texttt{NTL}\footnotemark[4],
\texttt{ATLAS}\footnotemark[4] and can be used by general
purpose computer algebra systems such as \texttt{Sage}\footnotemark[4] or \texttt{Maple}\footnotemark[4].

We describe in this paper a selection of ideas and
improvements that were recently introduced into the the design of \linbox
for the forthcoming 2.0 release.

\section{The lightweight founding scope allocation model}

The main objects that require memory allocation in \linbox are base field or
ring elements, vectors, matrices, and polynomials.
The memory management for all of these object types follows the same
rules, organized to maximize efficiency in time and space, and
consequently requiring some efforts by the programmer: the allocations
and deallocations must always occur in the founding scope.
In particular
no external garbage collection mechanism is used.
\subsection{Call-by-reference}

The input and output types of most functions are usually template
types, and can be either basic types, or complicated
objects. Consequently, passing arguments by value (copy) must be avoided as
much as possible. Every argument is passed as a reference, including
the return types. More precisely the return value of a function is
also the first argument, defined as a non const reference. 
\begin{verbatim}
Matrix& someFunction(Matrix& result, const XXX& args);
\end{verbatim}
This convention was already presented in \cite[\S 2.1]{jgd:2002:icms} for the
design of field and ring arithmetic. It does require a redefinition of the interface
for some \texttt{stl}-like operators, as discussed in
section~\ref{ssec:parallel}.
A consequence of the above convention is that the objects returned by
a function,
have to be declared and initialized (in particular, memory allocated, e.g. via constructors) before the
function call.
By enforcing this
practice, we require that the programmer keep 
\textit{the handle} on the
objects that he allocates until all uses of the object and its subobjects are completed. Moreover, he is responsible for object 
deallocation in the same 
scope where it was allocated. 
This restricts some convenient programming practices, but provides precise control of memory usage.
This is particularly important when large, memory filling, matrices
are in play.
It also allows to avoid the costs of garbage 
collection or reference counting.
Many \linbox objects involve a handle containing a reference to the free store.
Note that even though a function does not allocate the handle itself,
it is in certain cases still free to resize and thus reallocate the free store memory referenced.

\begin{paragraph}
{\em Dense Matrix allocations.}
The objects storing dense matrices require a special care
concerning their allocations. Dense matrices are represented as a one
dimensional array storing elements in the row major format:
\texttt{A[i,j] = *(A+i*n+j)}.  It is important to be 
able to define a sub-matrix as a view on such an array, without
allocating the data.
For this we propose to distinguish two classes: one for allocated (via
constructors) matrices and the other for sub-matrix views. The 
genericity of  the template mechanism or inheritance will allow to use
these two types in the same code, without duplication. This allows
also for an automatic decision about deallocation.
Other solutions includes reference counting and explicit "end of use"
functions.

 Thus a first approach is to define a dense matrix class with a
 boolean \texttt{\_alloc} 
 member, telling whether the matrix owns its data or whether it is a
 simple view 
 on some other matrix's data. The destructor deallocates the data only if
 \texttt{\_alloc} is \texttt{true}. This can be viewed as a simplified
 reference counting mechanism, where one assumes that the matrix
 initially allocated is always destructed after all of its
 sub-matrices. This convention is consistent with the previous
 consideration: the allocations and deallocations must always occur in
 the founding scope.

 To further improve the efficiency, an alternative is to distinguish
 two classes: 
 one for allocated matrices and the other for sub-matrix views. The
 genericity of 
  the template mechanism or inheritance will allow to use these two
  types in the same code, without duplication. Furthermore,
  thread-safety mechanism on the \texttt{\_alloc} 
 member are not required anymore.

Remark that in this founding scope model, neither the alloc variable nor a two
classes system is required.
The programmer should know whether a matrix is created as a sub-matrix
or as an allocating instance by what constructors or other initializers
she uses. Thus she knows which require care to deallocate in the same scope.
What she does not necessarily get is automatic decision about deletion
in the destructor, and would thus have to call an explicit "end of use"
function.
\end{paragraph}

\subsection{Rebind of coefficient domains}
\subsubsection{Mapping of data between domains}
\linbox makes use of the concept of rebinds for the mapping of data
structures between different coefficient domains.
For instance, in the context of the Chinese remainder algorithm, rebinds
allow to map a matrix over the integers of type, say,
(\texttt{DenseMatrix<PID\_Integer>}) to a modular matrix of type, say,
\\(\texttt{DenseMatrix<Modular<double> >}).

In \linbox, binder adaptors are enclosed
within many data structures and make use of a generic
converter, named \texttt{Hom} and found in \url{linbox/field/hom.h}.
\texttt{Hom} can generically use the \linbox domain's canonical
conversion methods methods \url{init} and \url{convert} from/to the \linbox
Integer type: \texttt{Domain1} $\rightarrow$ \texttt{Integer}
$\rightarrow$ \texttt{Domain2}. 
Moreover, when natural, efficient conversions exists between domains
(e.g. different representations of the same field or one ring embedded in another), generic \texttt{Hom} can be directly bypassed by a specialization~of~\texttt{Hom}.

\begin{paragraph}
{\em Rebind of dense matrices.}
We illustrate the founding scope allocation model with the use of rebind
functions adapted from the allocators in the STL, on dense matrices.
\begin{verbatim}
template <class Domain> class DenseMatrix {
  typedef DenseMatrix<Domain> Self_t;
  ... 
  template<class AnyDomain> struct rebind{ 
     typedef DenseMatrix <AnyDomain> other;
     operator ()(other& Ap, const Self_t& A, const AnyDomain& D){
       // Performs the modular conversion of A to Ap over D
       typename Self_t::ConstRawIterator A_iter;
       typename other::RawIterator Ap_iter;
       Hom<Field, _Tp1> hom(A. field(), F);
       for (A_iter = A. rawBegin(), Ap_iter = Ap.rawBegin();
            A_iter != A. rawEnd(); ++ A_iter, ++ Ap_iter)
         hom.image (*Ap_iter, *A_iter);
     } 
  };  
}
\end{verbatim}
According to the founding scope allocation model, the function
\texttt{operator()} in charge of the initialization of the matrix cannot 
allocate any memory. This has to be done at the level where the
rebind is called. This also requires a modification of the rebind
operator interface of the STL: the new object is passed by reference.

\end{paragraph}

\subsubsection{Rebind of handlers in the founding scope allocation model}
In the case of BlackBoxes (functions providing only a matrix-vector
product and not necessarily storing any data) the rebind mechanism
becomes more specific. We detail in this section the solution provided
in blackboxes which only store references to other blackboxes, such as
the \texttt{Compose}, \texttt{Transpose}, \texttt{Submatrix}, etc.

Indeed to rebind a blackbox containing only references one should
allocate a new memory zone and rebind the refered blackbox there.
The problem is that a caller, given a \texttt{BlackBox::rebind<Field2>::other}
type, does not necessarily know how to allocate for this object. 
The STL solution is to embed the allocator in each container. 
In \linbox, we propose another solution: the \texttt{other} not only
has different elements, but also can be of a different type.
For instance, the rebind \texttt{other} type of a blackbox containing
a reference will be the same kind of blackbox, but physically storing
the data (and thus owning it). 

For the different blackboxes defined in \linbox which use references,
we thus define a similar class called e.g. \texttt{TransposeOwner},
\texttt{ComposeOwner}, \texttt{SubmatrixOwner}, etc.
These classes store and own their data. Then it suffices for the
rebind sub-class of their reference equivalents to define its
\texttt{other} type to the associated \verb!*Owner! class. The example
of the \texttt{Compose} is given in figure \ref{fig:composeowner}.
\begin{figure}[htbp]
\begin{verbatim}
template <class _Blackbox1, class _Blackbox2> class Compose {
...
  template<typename _Tp1, typename _Tp2 = _Tp1> struct rebind {
    typedef ComposeOwner<
               typename Blackbox1::template rebind<_Tp1>::other,
               typename Blackbox2::template rebind<_Tp2>::other
    > other;
    ...
  };
  const Blackbox1 * _A_ptr;
  const Blackbox2 * _B_ptr;
  
};

template <class _Blackbox1, class _Blackbox2> class ComposeOwner {
...
  template<typename _Tp1, typename _Tp2 = _Tp1> struct rebind {
    typedef ComposeOwner<
               typename Blackbox1::template rebind<_Tp1>::other,
               typename Blackbox2::template rebind<_Tp2>::other
    > other;
    ...
  };
  template<typename _BBt1, typename _BBt2, typename Field>
  ComposeOwner (const Compose<_BBt1, _BBt2> &M, const Field& F)
    : _A_data(*(M.getLeftPtr()), F), _B_data(*(M.getRightPtr()), F) {
        typename Compose<_BBt1, _BBt2>::template rebind<Field>()(*this,M,F);
  }
  template<typename _BBt1, typename _BBt2, typename Field>
  ComposeOwner (const ComposeOwner<_BBt1, _BBt2> &M, const Field& F)
    : _A_data(M.getLeftData(), F), _B_data(M.getRightData(), F) {
        typename ComposeOwner<_BBt1, _BBt2>::template rebind<Field>()(*this,M,F);
  }

  Blackbox1 _A_data;
  Blackbox2 _B_data;
};
\end{verbatim}
\caption{Owner mechanism for the composed blackboxes}\label{fig:composeowner}
\end{figure}

Not this also fits well in the \linbox
founding scope allocation, since the \verb!*Owner! class will be
declared (and thus allocated) by the caller of the rebind in codes
similar to the following:
\begin{verbatim}
template<class BlackBox> void f(const BlackBox& A) {
...
typedef typename Blackbox::template rebind<Field2>::other FBlackbox;
// rebinds generically the BlackBox A to a BlackBox Ap 
// with a new Domain F2
// The container type of Ap might be different from the one of A
// this decision is made in the rebind type of A, 
// via the 'other' typedef
FBlackbox Ap(A, F2);
...
}
\end{verbatim}

Remark that for a submatrix of a class storing its elements (contrary
to a submatrix of a blackbox containing e.g. only references), a more
efficient rebind would only rebind the elements within the boundaries
of the submatrix. There we use a trait to decide wether the refered
blackbox is a storing component and in the latter case specialize
e.g. \texttt{Submatrix<DenseMatrix<Field1> >::rebind<Field2>::other}
to a simpler \texttt{DenseMatrix<Field2>} instead of using the
\verb!*Owner! mechanism.
\section{Software abstraction layer for parallelism}

Efficient parallel applications must take into consideration hardware
characteristics (number of cores, memory hierarchy, etc.). It is time
consuming or impossible for a single developer to 
program a high performance computer algebra application, with state of
the art algorithms, while exploiting all the available parallelism.  
In order to separate the domains of expertise we have designed a
software abstraction layer between computer algebra algorithms
and parallel implementations which may employ automatic dynamic scheduling.

\subsection{Parallel building blocks}\label{ssec:parallel}
Computer algebra algorithms have three main characteristics:
1) they are complex and require a deep knowledge of the problem in
  order to obtain the most efficient sequential algorithm;
2) they may be highly irregular. This enforces a runtime use of
  load balancing algorithms;
3) they are generic in the sense that they are usually designed
  to work over several algebraic domains.

  In the case of \linbox algorithms, we have decided to base our
  software abstraction, called {\em Parallel Building Blocks (PBB)},
  on the STL algorithms (Standard Template Like) principles.
  Indeed, C++ data structures in \linbox let us have random access
  iterators over containers which are naturally parallel. 
  We have already defined several STL-like algorithms and the list
  will be extended in the near future:\\
  {\bf for\_each, transform,
    accumulate\footnote{\url{www.sgi.com/tech/stl}}}: the PBB
  versions of these algorithms are similar to the STL versions 
  except that the involved operators (or function object classes), given as 
  parameters, are required to have their return value reference passed as the
  first parameter of the function. This is in accordance with the memory model 
  of \linbox. 
  The STL return-by-value semantic is not appropriate. 

  
  The fundamental idea of PBB is that at the computer algebra
  level, the parallelization of all the loops and more generally of all
  the STL-like algorithms will already enable good performance and
  easy switching among multiple implementations.
  Regarding performance, this parallelization of the inner loops of
  the underlying linear algebra is sufficient in many cases.
  Regarding implementations, this
  abstraction provides for programming independent of the
  parallel model with selection of the parallel environment
  depending on the target architecture.
  The parallel blocks can be implemented using many different parallel
  environments, such as
  OpenMP\footnote{\url{openmp.org}, \url{threadingbuildingblocks.org}}; 
  TBB\footnotemark[7] (Thread Building Blocks)
   or
  Kaapi~\cite{inproceedingsgautier.gbp_ktsrsf_07}; using
  both static and dynamic work-stealing
  schedulers~\cite{con-traore.trmgb_08}.
  The current implementations are built on OpenMP and Kaapi.

\subsection{\Accumulatewhile and early termination}
To bound the complexity of many linear algebra problems, one of the
key ideas is to use an accumulation with {\em early termination}.

For instance, this is used in Chinese Remaindering algorithms. The
computation is performed modulo a sequence of (co)prime numbers and
the result is built from a sequence of residues, {\em until} a
condition is satisfied~\cite{jgd:2010:crt}. 
The termination of the algorithm depends on the accumulated 
result.
  
In order to parallelize such algorithms, we proposed an extension of
the STL algorithms called \accumulatewhile.
The algorithm takes an array $v$ of length $N$, a unary
operator $f$ to be applied to each array entry and a specific binary
update operator/predicate for the accumulation.
This {\em accumulator} with a signature like \verb!bool accum(a, b)! behaves like an in place addition (\verb!a+=b!) and
returns \texttt{true} to indicate sufficiently many values are accumulated.
Let $S_k = \sum_{i=0,..,k} f( v[i])$ with 
$k \in \{0,N\}$.  The algorithm computes and returns $n \leq N$ and
$S_n$ such that one accumulation during the computation of $S_n$
returned \texttt{true} or $n = N$.  In intended use, we know any additional accumulation would also return \texttt{true}.

This algorithm will be used 
for the early termination Chinese remaindering algorithms of
\linbox. Though not yet using PBB and \accumulatewhile, a
sequential version and parallel versions with OpenMP and
Kaapi can be found in the \linbox distributions as 
\url{linbox/algorithms/cra-domain-*.h}.


\subsection{Memory contention and thread safe allocation}
Many computer algebra programs allocate dynamic memory for the
intermediate computations. Several experiments with \linbox
algorithms on multicore architectures have shown that these
allocations are quite often the bottleneck.
An analysis of the memory pattern and experiments with three well
known memory allocators 
(ptmalloc, Hoard and TCMalloc from Google Perf. Tools~\cite{tcmalloc})
have been conducted. The goal was to decide whether the parallel
building blocks model was suitable to high-performance exact linear
algebra. We used dynamic libraries to exchange allocators for the
experiments, but one can use them together in the \linbox library if
needed \cite[\S 7]{kaltofen:2005:memory}.
Preliminary experiments on early terminated Chinese remaindering,
not the easiest to parallelize, have demonstrated the advantage, in
our setting, of TCMalloc over the others~\cite{jgd:2010:crt}.
One of the main reasons for that fact is that our problems required
many temporary allocations. This fits precisely the thread safe caching
mechanism of TCMalloc.
\section{Automated Generic Separate compilation}
\linbox is developed with several kinds of genericity:
1) genericity with respect to the domain of the coefficients,
2) genericity with respect to the data structure of the matrices,
3) genericity with respect to the intermediate algorithms.
While this is efficient in terms of capabilities and code reusability, there is a combinatorial explosion of combinations.  Consider that each of 50 arithmetic domains may be combined with each of 50 matrix representations in each of 10 intermediate algorithm forms for a single problem as simple as matrix rank. This
lengthens the compilation time and generates large executable files.

For the management of code bloat \linbox has used an ``archetype
mechanism'' which enables, at the user's option, to switch to a
compilation against abstract classes \cite[\S 2.1]{jgd:2002:icms}.
However, this can reduce the efficiency of the library. Therefore, we propose
here a way to provide a generic separate compilation. This will not
deal with code bloat, but will reduce the compilation time while
preserving high performance.
This is useful for instance when the library is used with
unspecialized calls. This is largely the case for some interface
wrappers to other Computer algebra systems such as {\sc Sage} or {\sc Maple}.
Our idea is to automate the technique of
\cite{Erlingsson:1996:issac} which combines compile-time instantiation
and link-time instantiation, while using template instantiation
instead of void pointers.
The mechanism we propose is independent of the desired generic method,
the candidate
for separate compilation, and is explained in algorithm \ref{alg:sep}.
\begin{algorithm}[ht]
\caption{C++ Automatic separate compilation wrapping}\label{alg:sep}
\begin{algorithmic}[1]
\REQUIRE A generic function \texttt{func}.
\REQUIRE Template parameters \texttt{TParam} for separate
specialization/compilation of \texttt{func}.
\ENSURE A generic function calling
\texttt{func} with separately compiled instantiations.
\STATE Create a header and a body files ``func\_instantiate.hpp'' and ``func\_instantiate.cpp'';
\STATE Add a template function \texttt{func\_separate}, with the same
specification as \texttt{func}, to the header;
\STATE Its generic default implementation is a single line calling the
original function \texttt{func}.\\ \COMMENT{This enables to have a
  unified interface, even for non specialized class.}
\FOR{each separately compiled template parameter \texttt{TParam}}
  \STATE Add  a non template specification
  \texttt{funcTParam}, to the header file;
  \STATE Add the associated body with a
  single line returning the instantiation of
  \texttt{func} on a parameter of type \texttt{TParam}, to the body file;
  \STATE Add an inline specialization
  body of \texttt{func\_separate} on a parameter of type
  \texttt{TParam} with a single line returning \texttt{funcTParam}, to
  the header file; 
\ENDFOR
\STATE Compile the body file ``func\_instantiate.cpp''.
\end{algorithmic}
\end{algorithm}

\begin{figure}[ht]
\includegraphics[width=\textwidth]{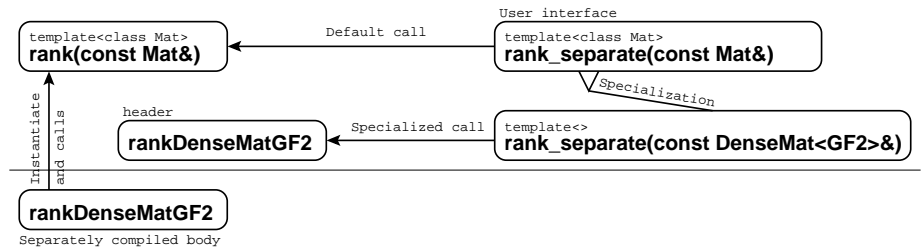}
\caption{Separate compilation of the rank}\label{fig:sep}
\end{figure}

This Algorithm is illustrated on figure \ref{fig:sep}, where
the function is the \texttt{rank} and the template parameter is a dense
matrix over $GF(2)$,
\texttt{DenseMatrix<GF2>}.

%
  Algorithm \ref{alg:sep} has been simplified for the
  sake of clarity. To enable a more user-friendly interface one can
  rename the original function and all its 
  original specializations \texttt{func\_original}; then rename also
  the new interface
 simply \texttt{func}. 
%
With the classical inline compiler optimizations, the overhead of
calling \texttt{func\_separate} is limited to single supplementary
function call. Indeed all the one line additional methods will be
automatically inlined, except, of course, the one calling the separately
compiled code.
If this overhead is too expensive, it suffices to enclose all the non generic specializations of
``func\_instantiate.hpp'' by a macro test. 
At compile time, the decision to separately
compile or not can be taken according to the definition of this
macro. 

We show in tables \ref{tab:compilation} and \ref{tab:compintel} the
gains in compilation time
obtained on two examples from \linbox: the \texttt{examples/\{rank,solve\}.C} algorithms. 
 Indeed, without any specification
 the code has to invoke several specializations depending on
 run-time discovered properties of the input. For instance
 \texttt{solve.C} requires 6 specializations for sparse
 matrices over the Integers or over a prime field, with a sparse
 elimination, or an iterative method, or a dense method, if the matrix
 is small\ldots
\begin{table}[ht]\center
\begin{tabular}{|l||r|r|r||r|r|r|}
\hline
file                      &  real time   &  user time   &  sys. time  &  real time   &  user time   &  sys. time \\
\hline
 & \multicolumn{3}{|c||}{Rank}& \multicolumn{3}{|c|}{Solve}\\
\hline
\texttt{instantiate.o} & 143.43s & 142.47s & 0.90s & 171.62s & 170.42s & 1.12s\\
\texttt{\{rank,solve\}.o} & \bf 18.58s & \bf 18.26s & \bf 0.30s & \bf 23.13s & \bf 22.80s & \bf 0.32s\\
\texttt{link} & 0.80s & 0.64s & 0.15s & 0.85s & 0.70s & 0.14s\\
\hline
\texttt{Sep. comp. total} & 162.81s & 161.37s & 1.35s & 195.60s & 193.92s & 1.58s\\
\hline
\texttt{Full comp.} & 162.02s & 160.47s & 1.21s & 191.47s & 189.52s & 1.40s\\
\hline
\hline
\texttt{speed-up} & 8.4 & 8.5 & 2.7 & 8.0 & 8.1 & 3.0s\\
\hline
\end{tabular} 
\caption{linbox/examples/\{rank,solve\}.C compilation time on an AMD
  Athlon 3600+, 1.9GHz, with gcc 4.5 -O2. \texttt{instantiate.o} contains to the separately compiled
  instantiations (e.g. densegf2rank in figure \ref{fig:sep});
  \texttt{\{rank,solve\}.o} contains to the user interface and generic
  implementation compilation; \texttt{link} corresponds to the linking
  of both \texttt{.o} and the library; \texttt{Full comp.} corresponds
  to the compilation without the separate
  mechanism.}\label{tab:compilation}
\end{table}

\begin{table}[ht]\center
\begin{tabular}{|l||r|r|r||r|r|r|}
\hline
file                      &  real time   &  user time   &  sys. time  &  real time   &  user time   &  sys. time \\
\hline
 & \multicolumn{3}{|c||}{Rank}& \multicolumn{3}{|c|}{Solve}\\
\hline
\texttt{instantiate.o} & 46.36s & 44.47s & 1.33s & 67.32s & 63.16s & 2.20s\\
\texttt{separate.o} & \bf 9.51s & \bf 9.13s & \bf 0.30s & \bf 9.88s & \bf 9.38s & \bf 0.30s\\
\texttt{separate} & 0.55s & 0.34s & 0.07s & 0.97s & 0.72s & 0.08s\\
\hline
\texttt{Sep. comp.} & 56.42s & 53.94s & 1.70s & 78.17s & 73.26s & 2.58s\\
\hline
\texttt{Full comp.} & 50.60s & 46.88s & 1.90s & 70.42s & 65.55s & 2.42s\\
\hline
\hline
\texttt{speed-up} & 5.0 & 5.0 & 5.1 & 6.5 & 6.5 & 6.4\\
\hline
\end{tabular} 
\caption{linbox/examples/\{rank,solve\}.C compilation time on an intel
  Xeon E5345, 2.33GHz, with icc-11.1 -O2.}\label{tab:compintel}
\end{table}

\section*{Acknowledgment}
We thank the \linbox group and especially Brice Boyer, Pascal Giorgi,
Erich Kaltofen, Dan Roche, Brian Youse for many useful discussions 
in particular during the recent \linbox developer meetings in
Delaware and Dublin.
%
\bibliographystyle{abbrv}
\bibliography{icms} 

\begin{thebibliography}{1}

\bibitem{CEKSTV:2002:EP}
L.~Chen, W.~Eberly, E.~Kaltofen, B.~D. Saunders, W.~J. Turner, and G.~Villard.
\newblock {Efficient matrix preconditioners for black box linear algebra}.
\newblock {\em Linear Algebra and its Applications}, 343-344:119--146, 2002.

\bibitem{jgd:2002:icms}
J.-G. Dumas, T.~Gautier, M.~Giesbrecht, P.~Giorgi, B.~Hovinen, E.~Kaltofen,
  B.~D. Saunders, W.~J. Turner, and G.~Villard.
\newblock {LinBox}: A generic library for exact linear algebra.
\newblock In A.~M. Cohen, X.-S. Gao, and N.~Takayama, editors, {\em Proceedings
  of the 2002 International Congress of Mathematical Software, Beijing, China},
  pages 40--50. World Scientific Pub., Aug. 2002.

\bibitem{jgd:2010:crt}
J.-G. Dumas, T.~Gautier, and J.-L. Roch.
\newblock Generic design of chinese remaindering schemes.
\newblock In M.~Moreno-Maza and J.-L. Roch, editors, {\em PASCO 2010}.
  Universit\'e de Grenoble, France, July 2010.

\bibitem{DGP:2008:dlaff}
J.-G. Dumas, P.~Giorgi, and C.~Pernet.
\newblock Dense linear algebra over word-size prime fields: the fflas and
  ffpack packages.
\newblock {\em ACM Trans. Math. Softw.}, 35(3):1--42, 2008.

\bibitem{Erlingsson:1996:issac}
U.~Erlingsson, E.~Kaltofen, and D.~Musser.
\newblock Generic {Gram}-{Schmidt} orthogonalization by exact division.
\newblock In {\em ISSAC'1996}, pages 275--282, July 1996.

\bibitem{inproceedingsgautier.gbp_ktsrsf_07}
T.~Gautier, X.~Besseron, and L.~Pigeon.
\newblock {KAAPI:} a thread scheduling runtime system for data flow
  computations on cluster of multi-processors.
\newblock In {\em PASCO'07}, pages 15--23, 2007.

\bibitem{tcmalloc}
S.~Ghemawat and P.~Menage.
\newblock Tcmalloc: Thread-caching malloc.
\newblock \url{http://goog-perftools.sourceforge.net/doc/tcmalloc.html}.

\bibitem{kaltofen:2005:memory}
E.~Kaltofen, D.~Morozov, and G.~Yuhasz.
\newblock Generic matrix multiplication and memory management in \linbox.
\newblock In {\em {ISSAC}'2005}, pages 216--223, July 2005.

\bibitem{con-traore.trmgb_08}
D.~Traore, J.~L. Roch, N.~Maillard, T.~Gautier, and J.~Bernard.
\newblock Deque-free work-optimal parallel {STL} algorithms.
\newblock In {\em EUROPAR 2008}, pages 887--897, Las Palmas, Spain, Aug. 2008.

\end{thebibliography}
%
\end{document}